\documentclass[floatfix,aps,prd,showpacs,amsmath,nofootinbib,preprintnumbers,twocolumn]{revtex4}
\usepackage{graphicx}
\def\beqan{\begin{eqnarray*}}
\def\eeqan{\end{eqnarray*}}
\newcommand{\beqa}{\begin{eqnarray}}
\newcommand{\eeqa}{\end{eqnarray}}
\newcommand{\beq}{\begin{equation}}
\newcommand{\eeq}{\end{equation}}
\newcommand{\bal}{\begin{align}}
\newcommand{\eal}{\end{align}}

\def\bear{\begin{array}}
\def\enar{\end{array}}

\def\mphi{m_{\phi}}
\def\mchi{m_{\chi}}
\def\mi{m_I}
\def\h2{\frac{\mi^2}{M_{p}}}
\def\Mp{M_{p}}

\def\bear{\begin{array}}
\def\enar{\end{array}}

\begin{document}

\preprint{\vbox{ \hbox{} \hbox{hep-ph/0703319} \hbox{March 2007} }}


\title{A note on the moduli-induced gravitino problem.}

\author{Natalia Shuhmaher}
\email[Email: ]{ natalia.shuhmaher@physics.unige.ch}

\affiliation{Department of Physics, McGill University, Montr\'eal,
QC, H3A 2T8, CANADA \\ D\'{e}partement de Physique Th\'{e}orique,
Universit\'{e} de Gen\'{e}ve, 24 Ernest Ansermet, 1211 Gen\'{e}ve 4,
Switzerland}

\begin{abstract} 

The cosmological moduli problem has been recently reconsidered.
Papers~\cite{Endo:2006zj,Nakamura:2006uc} show that even heavy
moduli ($m_\phi > 10^5$ GeV) can be a problem for cosmology if a
branching ratio of the modulus into gravitini is large. In this
paper, we discuss the tachyonic decay of moduli into the Standard
Model's degrees of freedom, e.g. Higgs particles, as a resolution to
the moduli-induced gravitino problem. Rough estimates on model
dependent parameters set a lower bound on the allowed moduli at
around $10^8 \sim 10^9$ GeV.

\end{abstract}

\pacs{98.80.Cq.}

\maketitle

\section{Introduction}

The cosmological moduli problem is a disease of many
supersymmetry/supergravity
theories~\cite{Coughlan,Ellis2,Carlos,Banks:1993en}. Many
supersymmetry/supergravity theories contain fields which have flat
potentials in the supersymetric limit and only Planck suppressed
couplings to Standard Model (SM) particles. We generically call them
moduli. The cosmological moduli problem arises whenever decays of
moduli are in conflict with cosmological observations. Masses of
moduli depend on the type of supersymmetry breaking. Moduli much
lighter than the Hubble scale during inflation acquire a vacuum
expectation value (VEV) of order the Planck
scale~\cite{Linde,FordVil} and even exceed it if the mass of moduli
is not sufficiently high~\cite{Starobinsky:1986fx,Podolsky:2002qv}.
In the last case, the modulus field can become an inflaton. Later
on, a large abundance of moduli threatens to overclose the Universe
or jeopardize the processes of nucleosynthesis. Several solutions of
the moduli problem have been suggested, see
e.g.~\cite{Lazarides:1985bj,Lazarides:1985ja,Dine:1995uk,Shuhmaher:2005mf}.

The cosmological moduli problem is automatically avoided in heavy
moduli scenarios. A widely used estimate for the perturbative decay
rate $\Gamma_{all}$ of moduli is \beq \Gamma_{all} \, ~\sim \, {1
\over 4 \pi} {{\mphi^3} \over {\Mp^2}} \, . \label{decAll}\eeq where
$\phi $ is the modulus field and $m_\phi$ is the modulus mass.
Moduli decay once the Hubble rate is of the order of $\Gamma_{all}$.
Therefore, moduli of mass below 100 TeV decay near or after the time
of nucleosynthesis, when the universe is nearly $1$ second old. If
the mass is above $100$ TeV then the moduli decay before the time of
Big Bang Nucleosynthesis (BBN). Examples of scenarios with heavy
moduli
exist~\cite{Moroi:1994rs,Randall:1994fr,Kawasaki:1995cy,Moroi:1999zb,Kachru:2003aw}.

The heavy moduli scenario as a solution of the cosmological moduli
problem has recently been reconsidered starting with the
papers~\cite{Endo:2006zj,Nakamura:2006uc}. It was shown that the
decay of moduli into gravitinos is unsuppressed (for an opposite
example see~\cite{Dine}). The part of the {\cal L}agrangian
describing the gravitino-modulus couplings is \beqa e^{-1} {\cal L}
&=& -{1 \over 8} \epsilon^{\mu \nu \rho \sigma} (G_\phi
\partial_\rho \phi + G_{\phi^\dagger} \partial_\rho
\phi^\dagger) \bar{\psi}_\mu \gamma_\nu \psi_\sigma \\ & & -{1 \over
8} e^{G/2} (G_\phi \phi + G_{\phi^\dagger} \phi^\dagger)
\bar{\psi}_\mu [\gamma^\mu, \gamma^\nu]  \psi_\nu \eeqa where
$\psi_\mu$ stands for the gravitino and $G_\phi$ is a non vanishing
dimensionless auxiliary field  with $G = K/M^2_p +
\ln(|W|^2/M^6_p)$. The subscript $i$ denotes the derivative with
respect to the field $i$. $K$ and $W$ are K\"{a}hler potential and
superpotential respectively. Based on these coupling, the
perturbative decay rate of moduli into gravitinos is \beq
\Gamma_{3/2} \equiv \Gamma(\phi \rightarrow 2 \psi_{3/2}) \approx
{|G_\phi|^2 \over 288 \pi} {m^5_\phi \over m^2_{3/2} M^2_p}
\label{decGrav}\, . \eeq

The auxiliary field of the modulus, $G_\phi$, in general, can be
small to suppress $\Gamma_{3/2}$ to the total decay rate
$\Gamma_{all}$~(\ref{decAll}). However, suppressed $G_\phi$ is not a
typical case, e.g. in the framework of the $4D$ supergravity $G_\phi
\gtrsim m_{3/2}/m_\phi$. Performing elaborate calculations, authors
of~\cite{Endo:2006zj,Nakamura:2006uc} have shown that the typical
branching ratio $Br(\phi \rightarrow 2 \psi_{3/2}) \sim {\cal
O}(0.01-1)$. The large branching ratio of heavy moduli into
gravitinos causes gravitino overproduction. Hence, even having a
modulus mass above 100 TeV does not resolve the cosmological moduli
problem. A detailed re-analysis of the cosmological moduli problem
taking into account constraints on gravitino overproduction pushes
up the gravitino mass above $10^5 - 10^6$
GeV~\cite{Nakamura:2006uc}. This is the moduli-induced gravitino
problem.

The previously published literature on the moduli-induced gravitino
problem does not include nonperturbative decay channels. We propose
a solution of the moduli-induced gravitino problem by having most of
the moduli energy decay into the SM degrees of freedom through a
tachyonic decay into a boson pair, e.g. Higgs. The decay process
{\it moduli $->$ bosons} is rapid and occurs before moduli start to
perturbatively decay into gravitinos. The scheme allows to find a
range of scalar moduli masses ($m_\phi > 10^8 \sim 10^9$ GeV) which
does not suffer from the moduli-induced gravitino problem. Making
use of conservative approximations, we find a range of masses with
no overproduction of gravitinos.

\section{Basic Idea}

The general idea can be introduced in the following way. As was
mentioned previously, moduli have only Planck suppressed couplings
to other fields and during inflation obtain a VEV of the order of
the Planck scale. After inflation, the modulus field slowly rolls
preserving its energy. When the Hubble parameter reaches the value
of $m_\phi$, the modulus field starts to oscillate. In the
following, we assume that moduli have a trilinear coupling to a
scalar field $\chi$, \beq \phi \chi^2 \label{triliniar} \, .\eeq The
effective potential, $V(\phi,\chi)$ is \beq V(\phi,\chi) =
\frac{1}{2} \mphi^2 \phi^2 + \frac{1}{2} \mchi^2 \chi^2 + {1 \over
2} \frac{\alpha}{M_{p}} m^2_\phi \phi \chi^2 + {1 \over 4}\lambda
\chi^4  \, . \label{potenPchi} \eeq The equation of motion for
$\chi$ field with switched off the expansion of space is \beq \ddot{
\chi}_k  + \left( {k^2} + m^2_{eff} \right) \chi_k \, = \, 0 \, .
\label{eq:Mattype} \eeq where \beq m^2_{eff} = \mchi^2 + \lambda
\chi^2 + \frac{\alpha}{M_p} m^2_\phi \phi \, , \eeq The oscillations
of the field $\phi$ induce a negative mass for the field $\chi$. The
modes of the field $\chi$ with $k < \sqrt{-m^2_{eff}}$ are excited,
\beq \chi_k \propto e^{\sqrt{-m^2_{eff}- k^2}t}\label{modes} \eeq
and the energy is transferred from the oscillating $\phi$ into
excitations of $\chi$ in a preheating-like process. The process has
a name of tachyonic resonance or tachyonic (p)reheating and is
widely discussed in the literature starting
with~\cite{TB,Greene:1997ge,Felder:2000hj}, in particular, the
implementation of tachyonic resonance in the context of the
resolution of the moduli problem is discussed
in~\cite{Shuhmaher:2005mf}. Thus, we see that for a certain range of
parameters, the energy density stored in the moduli
nonperturbatively transfers into excitations of $\chi$ field much
before moduli perturbatively decay into gravitinos. The couplings of
$\chi$ to Standard Model particles are assumed to be unsuppressed
and, as a result, the decay rate of $\chi$ is much larger than $1$
sec$^{-1}$. Thus, the modulus energy is converted into radiation
much before the time of BBN.

To study the stability of the potential~(\ref{potenPchi}), we find
the minimum of the $V(\phi,\chi)$ in the $\phi$ direction which
occurs for \beq \phi = -{1 \over 2} \frac{\alpha}{M_p} \chi^2 \, .
\label{minphi} \eeq Substituting~(\ref{minphi}) into $V(\phi,\chi)$
leads to \beq V(\phi,\chi) = - {1 \over 4} \left( {1 \over 2}
\frac{\alpha^2}{M^2_p}m^2_\phi - \lambda \right) \chi^4 + {1 \over
2} m^2_\chi \chi^2 \, ,\eeq and, we see that the effective potential
is unstable for \beq {1 \over 2} \frac{\alpha^2}{M^2_p}m^2_\phi >
\lambda \label{condLambda}\, .\eeq Thus, the presence of additional
terms with Planck suppressed couplings is important to stabilize the
potential~(\ref{potenPchi}) at large values of the fields.

The efficiency of the tachyonic resonance must be carefully checked
against the effects of dilution due to the expansion of space. For
the tachyonic resonance to be effective, the growth of the mode
$k$~(\ref{modes}) shall dominate the dilution due to the expansion
of space. The appropriate condition would be \beq \sqrt{-m^2_{eff}-
k^2}
> H \, \eeq or \beq \frac{\alpha}{M_p} m^2_\phi \Phi > {m^2_\phi
\Phi^2 \over M^2_p} \label{cond_res0} \eeq where $\Phi$ is the
amplitude of the $\phi$ field. In the last step, we replaced H with
the appropriate contribution from the modulus field. Further, we
make an assumption that the energy density of the modulus is the
significant component of the total energy density. If this is not
the case, the moduli-induced gravitino problem disappears. The
reason is that the produced gravitino represent only small portion
of the total energy density. The condition~(\ref{cond_res0}) is
fulfilled once \beq \alpha M_p
> \Phi \, . \label{cond_res}\eeq  At the onset of oscillations $\Phi
< M_p$, thus for $\alpha \geq 1$ we can neglect the expansion of
space in our analysis.

In addition to the growing mode~(\ref{modes}), there is also the
decaying mode \beq \chi_k \propto e^{-\sqrt{-m^2_{eff}-
k^2}t}\label{modes1} \, .\eeq The decaying mode causes inference
terms and may put further restrictions on the region of
applicability of the tachyonic resonance. The
equation~(\ref{eq:Mattype}) can take the form of the well known
Mathieu equation (see e.g.~\cite{Mathieu}). In fact as it can be
seen from the instability chart of the Mathieu equation, the
resonant production is terminated as soon as $q \equiv \alpha
\Phi/M_p \leq 1/2$; hence we are interested only in cases with
$\alpha \gg 1$. Models where $\alpha$ has to be smaller than 1 can
be of interest if many trilinear interactions enhance the resonant
effect. The condition on $\alpha$ is the same as in~(\ref{cond_res})
which means that the resonance production is efficient once the
change in the scale factor is negligible.

Tachyonic preheating in the parameter range corresponding to large
$\alpha$ was extensively studied in~\cite{Dufaux:2006ee}. The
authors have shown that trilinear terms lead to faster re-scattering
and thermalization. As a bonus, trilinear terms allow complete decay
of the moduli. In addition to positive effects, enhanced resonance
and fast subsequent thermalization may enlarge the reheating
temperature beyond the allowed region which threatens to overproduce
gravitinos through re-scattering processes~\cite{rescattering}.

The trilinear interaction term~(\ref{triliniar}) may arise, for
example, from the non-renormilizable term in the K\"{a}hler
potential \footnote{Here we provide only one example of the origin
of trilinear terms. Large $\alpha$ might require other
interactions.} \beq  \label{interaction1}  {\cal L}_H = \int d^4
\theta \frac{\lambda_H}{M_{p}} \phi H^*_u H^*_d +h.c. \eeq where
$H_u$ and $H_d$ are up-type and down-type Higgs supermultiplets or
corresponding scalar fields, respectively. The $\phi$ field is the
moduli supermultiplet and, in the following, its scalar part. After
integrating out the superspace coordinates, we obtain \beqa
\label{interaction1} {\cal L}_H &=& \frac{\lambda}{M_p}\left( D_\mu
D^\mu \phi H^*_u H^*_d \right. \\
& & \left. + F_\phi H^*_u F^*_d + F_\phi H^*_d F^*_u + c.c. + \cdots
\right) \nonumber \eeqa where $F_i = -M^2_p e^{G/2} (G^{-1})^i_j
G_j$ is the auxiliary field of the $i$'th supermultiplet, $D_\mu$ is
the covariant derivative. The process of energy transfer described
above makes use of on-shell degrees of freedom. Hence, we make use
of the equation of motion for the $\phi$ field to replace $D_\mu
D^\mu \phi$ with $m^2_\phi \phi$. As a result, the following
interaction term is a part of the {\cal L}agrangian: \beq {\cal L}_H
\supset \frac{\lambda}{M_{p}} m^2_\phi \phi H^*_u H^*_d + h.c.
\label{interactHiggs}\eeq In the low energy effective Lagrangian,
the term~(\ref{interactHiggs}) is responsible for the
interaction~(\ref{triliniar}), where $\chi$ is the neutral scalar
component of the lightest Higgs field in the mass basis.

\section{Estimates}

In the following we would like to estimate the region of moduli mass
for which the moduli-induced gravitino problem is resolved. Another
glance at the equation of motion of the $\chi$ field \beq \nonumber
\ddot{\chi}_k  + \left( k^2 +\mchi^2 + \lambda \chi^2 +
\frac{\alpha}{M_p} m^2_\phi \phi \right) \chi_k \, = \, 0 \, ,
\label{chi} \eeq reveals that the tachyonic process is more
effective for larger masses of the moduli. We assume that the
tachyonic resonance works as long as $m^2_{eff}$ can obtain negative
values, \beq \frac{m^2_\chi}{m^2_\phi} < \alpha \frac{\Phi}{M_p}
\label{effcon}\, . \eeq All the energy converted into excitations of
the $\chi$ field afterwards is transferred to SM degrees of freedom.
Further, since $Br_{3/2} = {\cal O}(0.01 \sim 1)$ we assume that
once the bound~(\ref{effcon}) is violated all the energy is
transferred to gravitinos. The above assumptions allow us to
estimate the gravitino abundance neglecting the effect of the
expansion of space. At the end, we insert the known bounds on the
gravitino abundance and derive the lower bound on the gravitino
mass.

We distinguish between two cases at the onset of moduli field
oscillations: in the first case, the universe is supercooled and
$\langle\chi^2\rangle \sim 0$; or, in the second case, the universe
is dominated by radiation and $\langle\chi^2\rangle \sim T^2 =
\sqrt{m_\phi M_p}$. The universe is supercooled if oscillations of
the moduli were preceded by an inflationary period, and the energy
is still stored in the oscillations of an inflaton, or if the
modulus itself is the inflaton
(see~\cite{Kawasaki:2006gs,Kawasaki:2006hm} for discussions on the
moduli-induced gravitino problem in this case). In this paper, we
primary concentrate on the first case. In this case, we omit the
self interaction term to obtain order of magnitude estimates for the
bound on the allowed moduli mass.

While the tachyonic resonance is in effect, the energy density in
$\phi$ is transferred to $\chi$ particles and then to radiation.
Neglecting the expansion of space, \beq \rho_{rad} = m^2_\phi M^2_p
\label{expansion} \eeq The tachyonic resonance ends as soon as
$\Phi$ reaches the value \beq \Phi_{min} = \frac{m^2_\chi}{m^2_\phi}
\frac{M_p}{\alpha} \, . \label{PhiMin} \eeq At this point, the
remaining energy density in the moduli is \beq m^2_\phi \Phi^2_{min}
= \frac{m^4_\chi M^2_p}{\alpha^2 m^2_\phi} \, \equiv \rho_{3/2} \, .
\eeq

The energy density stored in the gravitino, $\rho_{3/2}$, allows us
to determine the gravitino abundance. \beqa m_{3/2} Y_{3/2} &\equiv& m_{3/2} \frac{n_{3/2}}{s} \\
&=& \frac{\rho_{3/2}}{s} \\ &=& \frac{m^4_\chi M^2_p}{\alpha^2
m^2_\phi s} \eeqa where $Y_{3/2}$ is the gravitino yield, $n_{3/2}$
is the number density of gravitino particles and $s$ is the entropy
of the ultra-relativistic particles. \beq s=\frac{\rho+p}{T_R} =
\frac{4}{3} \frac{\rho_{rad}}{T_R} \approx (m_\phi M_p)^{3/2} \, ,
\label{entropy} \eeq where $T_R$ is the reheating temperature
(temperature of ultra-relativistic plasma at the moment it reaches
thermal equilibrium). While the actual reheating temperature depends
on the thermalization processes, the upper bound is \beq T_R <
\sqrt{m_\phi \Phi_{in}} \leq \sqrt{m_\phi M_p} \eeq where
$\Phi_{in}$ is the amplitude of the field $\phi$ at the onset of
oscillations. Since we have neglected the expansion of space
throughout the calculations, we have plugged $T_R = \sqrt{m_\phi
M_p}$ to obtain the last equality in~(\ref{entropy}).

The gravitino abundance is severely constrained in order not to
jeopardize the success of BBN or from the danger of overproducing of
lightest supersymmetric particles. The most stringent constraint
comes from the overproduction of
$^3He$~\cite{Kawasaki:2004yh,Kawasaki:2004qu} which yields \beq
m_{3/2} Y_{3/2} < O(10^{-14} \sim 10^{-11}) \mbox{ GeV} \, .
\label{gravAbun} \eeq The limit~(\ref{gravAbun}) is equivalent to
\beqa m_{3/2} Y_{3/2} &=& \frac{m^4_\chi }{\alpha^2 m^4_\phi } T_R
\\ &=& \frac{3}{4} \frac{m^4_\chi }{\alpha^2 m^4_\phi }
\sqrt{m_\phi M_p} \nonumber \\ &<& O(10^{-14} \sim 10^{-11}) \mbox{
GeV} \nonumber \eeqa where we have inserted the expression for
$s$~(\ref{entropy}). Making further assumptions: $\alpha \sim O(1)$,
$m_\chi \approx 100$ GeV, the moduli is safe from the overproduction
of gravitinos in direct decay if \beq 10^8 \sim 10^9 \mbox{ GeV}
\leq m_\phi \label{bound}\, .\eeq The lower bound~(\ref{bound}) is
the main result of the paper.

In the second case, when the field $\chi$ is a part of the thermal
bath and the contribution of the self interaction term to the
effective mass can be large, we have \beqa m^2_{eff} &=& \mchi^2 +
\lambda \langle \chi^2 \rangle + \frac{\alpha}{M_p} m^2_\phi \phi
\nonumber
\\ &=& \mchi^2 + \lambda T^2 + \frac{\alpha}{M_p} m^2_\phi \phi \,
\label{effmasschi} ,\eeqa  where we have used the Hartree
approximation to go from the first to the second line. The large
$\lambda T^2$ term threatens to prevent the tachyonic resonance from
occurring. Particulary, if, at the onset of oscillations, the
condition \beq 1 < \frac{\alpha}{\lambda} \frac{m_\phi}{M_p} \eeq is
not satisfied, the effective mass~(\ref{effmasschi}) is positive. In
an expanding moduli-dominated universe, the temperature redshifts as
\beq T^2=m_\phi M_p \left(\frac{\Phi}{M_p}\right)^{4/3}
\label{temp}\eeq Hence, $m^2_{eff}$ remains positive during
oscillations of the $\phi$ if \beq 1 > \frac{\alpha^3}{\lambda^3}
\frac{m_\phi}{M_p} \, \eeq where we have inserted $ \Phi_f =
{m^2_\phi \over M_p}$ - the value of $\Phi$ at the time of
perturbative decay~(\ref{decAll}). In the case $\mchi^2 > \lambda
T^2$, the estimates on moduli mass reduce
to~(\ref{expansion}-\ref{bound}).

The decay of moduli dilutes the pre-existing abundance of
gravitinos. Let us denote the initial gravitino yield by $Y_{3/2}$.
The entropy produced in the decay of moduli into radiation $s_n
\propto T^3_n$, hence, the new gravitino yield is \beqa Y^n_{3/2}&=&
\frac{n_{3/2}}{s_f+s_n} Y^n_{3/2} \approx \frac{Y_{3/2} s_f}{s_n} =
\frac{Y_{3/2} s_f}{s_n} \nonumber \\ &=& \frac{T^3_f}{T^3_n} Y_{3/2}
\, . \eeqa where $s_f$ and $T_f$ stands for the values of the
preexisting entropy and temperature of radiation at $\Gamma_{all} =
H$. Making use of~(\ref{temp}), we deduce \beq Y^n_{3/2}=
\frac{m_\phi}{M_p} Y_{3/2} \eeq

\section{Conclusions}

In this paper, we have discussed the influence of the tachyonic
resonance on the moduli-induced gravitino problem. We primarily have
discussed the case when $\chi$ is not a part of the thermal bath at
the onset of oscillations of the modulus field which is a main
contributor to the total energy density. In this case, the rough
estimates shows that moduli masses above $10^8 \sim 10^9$ are free
from overproduction of gravitinos in direct decay of moduli. The
estimates omit several model dependent points which may either
enhance or diminish the influence of the resonance. In particular,
in the process of calculations we did not take into account the
expansion of space. In the case when $\chi$ is a part of the thermal
bath at the onset of the oscillations of $\phi$, we have found that
the tachyonic resonance is less likely to work. In any case, even if
the tachyonic resonance is inefficient, the decay of moduli dilutes
the preexisting abundance of gravitinos. If the energy density of
the moduli is sufficiently subdominant to the total energy density,
the moduli-induced gravitino problem disappears. The reason is that
the produced gravitino represent only small portion of the total
energy density.

\acknowledgments We wish to thank Jean Dufaux, Masahiro Ibe and Lev
Kofman for useful discussions and to Alessio Notari for
proofreading. We are grateful to Robert Brandenberger (RB) for many
comments in the course of the project and proofreading our
manuscript. We would like to acknowledge support from a Carl
Reinhardt McGill Major Fellowship. This research is also supported
by an NSERC Discovery Grant to RB.



\end{document}